\documentclass[twocolumn,showpacs,preprintnumbers,amsmath,amssymb]{revtex4}
\usepackage{graphicx}
\usepackage{dcolumn}
\usepackage{bm}

\begin{document}
\preprint{to be submitted to PRD}

\title{Spherical Gravitational Wave Detectors:\\an approach for real time data analysis}

\author{C\'esar Augusto Costa}
\email{cesar@das.inpe.br}
\author{Odylio Denys de Aguiar}%
\email{odylio@das.inpe.br}
\affiliation{%
Instituto Nacional de Pesquisas Espaciais - Divis\~ao de Astrof\'isica\\
Av. dos Astronautas 1758. S\~ao Jos\'e dos Campos - SP - Brazil\\
}%
\date{\today}

\begin{abstract}
We have developed a full model to simulate spherical detectors where all main sources of noise are considered. We have built a computer code for determining the source direction and the wave polarization (solution of the inverse problem) in real time acquisition. The digital filter used is a simple bandpass filter. The ``data'' used for testing our code was simulated, which considers both the source signal and detector noise. The detector noise includes the antenna thermal, back action, phase noise, series noise and thermal from transducer coupled masses. The simulated noise takes into account all these noises and the antenna-transducers coupling. The detector transfer function was calculated for a spherical antenna with six two-mode parametric transducers. From the results we determined that spherical detectors are able to locate an astrophysical source in the sky in real time  as long as the signal-to-noise ratio for a burst signal is equal to or higher than 3.
\end{abstract}

\maketitle

\section{\label{sec:level1} Introduction}

Spherical gravitational wave detectors seem like an excellent low cost alternative for GW detection. They have  omnidirectional sensitivity and can give us some important information about an incident gravitational wave: its direction and tensorial components (polarization).

Moreover a sphere can be thought as 5 detectors in a single instrument. The cross-correlation between normal modes reduces noise effects and increases detection possibility.

Two of these instruments are being developed: the Brazilian Mario SCHENBERG and the Dutch MiniGRAIL   \cite{dewaard2006} and a third one is planned: the Sfera antenna \cite{fafone2006}. Both detectors have similar features and their data analysis approach can be used in the same way.

Robert Forward was the first to suggest a sphere as the antenna element in the early 1970's \cite{forward1971}. Ashby and Dreitlein studied the reception of GWs by an elastic self-gravitating spherical antenna and found a set of equations to treat such problem \cite{ashby1975}. Following up this idea Wagoner and Paik found the lowest eigenvalues for the monopole and quadrupole modes of a uniform elastic sphere and calculated its cross-section \cite{wagoner1977}.

In the 1990's, Johnson and Merkowitz studied the antenna-transducer coupling problem and found an optimum configuration, which minimizes the number of transducers while keeping them in a symmetric distribution on the antenna: the truncated icosahedron (TI) configuration \cite{johnson1993,merkowitz1997}. Magalh\~aes and collaborators showed that distributions with more resonators can also have interesting symmetric properties \cite{magalhaes1995,magalhaes1997a}. Lobo also suggested a more elaborated symmetric distribution \cite{lobo2000}.

\begin{figure}[!ht]
\centering
\includegraphics[width=8.4cm]{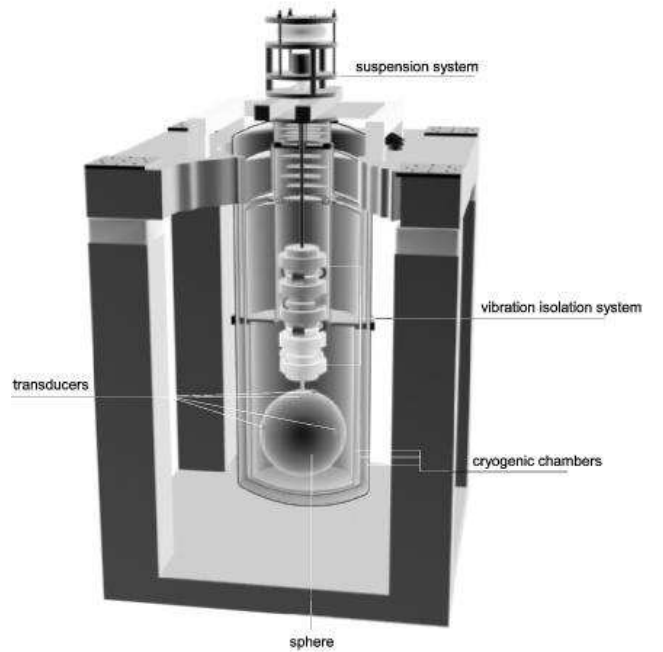}
\caption{\label{fig:schenberg}Schematic diagram of the Brazilian spherical GW detector Schenberg.}
\end{figure}
For more than a decade financial and technological difficulties had stalled the construction of spherical GW detectors but nowadays some research groups are working on this. One of these is the Brazilian Mario SCHENBERG detector (see Figure \ref{fig:schenberg}). The SCHENBERG has a $65\,\rm{cm}$-diameter and $\sim 1150\,\rm{kg}$ copper alloy $\rm{[Cu(94\%)Al(6\%)]}$ spherical antenna which is suspended by vibrational isolators. It will be kept under temperatures below $0.1\,\rm{K}$ inside cryogenics chambers \cite{grupo3}. These features are shared with the Dutch MiniGRAIL detector. The main difference between both is that the Brazilian detector will use parametric transducers while the Dutch data will be taken from capacitive transducers.

MiniGRAIL is ahead in the development. It has already started commissioned operations and soon will start scientific runs. Even though SCHENBERG is only now ready to start a commissioned phase at $4.2\,\rm{K}$, it has a data acquisition system and a data analysis software almost ready for taking scientific data.  We have developed a model based on harmonic oscillators to simulate their behavior. The development of software, which outputs simulated data, allowed us to create a methodology for initial data treatment. These procedures are presented in this paper along with how we can obtain information about GW sources directly from the detector data.

\section{The Model and the Simulated Data}

In order to develop a model that represents the response of a spherical detector to gravitational wave signals we created a set of displacement equations for the antenna plus resonator system. These equations (based on the linear elastic theory) are ruled by physical parameters such as element masses, spring constants, dumping coefficients and all coupling effects from electronic devices. The model considers the dynamics of a spherical antenna coupled to a set of electro-mechanical transducers and supply voltage pre-amplifiers readouts. In other words the model output is equivalent to the data which will be taken from the detector (voltage measurements which generate a time series).

The system dynamics can be expressed as a group of equations which represents coupled harmonic oscillators and is represented (already normalized) by
\begin{equation}\label{eq:moveq0}
\mathbf{\ddot{w}}(t)+
\mathbf{H}\mathbf{\dot{w}}(t)
+\mathbf{K}\mathbf{w}(t)=
\mathbf{M}^{-1}\mathbf{W}\mathbf{F}(t),
\end{equation}
where $\mathbf{M}$, $\mathbf{H}$ and $\mathbf{K}$ are respectively: the mass, dumping and spring matrixes. $\mathbf{w}(t)$ is the displacement vector and $\mathbf{F}(t)$ is the forces vector. $\mathbf{W}$ denotes the forces transfered to the system. This kind of equation is easily solved when one uses its orthonormal form in frequency domain \cite{merkowitz1996}. The solution is the following
\begin{equation}\label{eq:solmoveq}
\mathbf{\tilde{w}}(\omega)=\mathbf{U}\mathbf{\tilde{J}}(\omega)\mathbf{U}^\dag\mathbf{M}^{-1}\mathbf{W}\mathbf{\tilde{F}}(\omega),
\end{equation}
where $\mathbf{U}$ is the eigenvectors matrix and $\mathbf{U}^\dag$ is its hermitian conjugated. The system response function is represented by the matrix  ${\mathbf{\tilde{J}}^{-1}(\omega)\equiv-\omega^2\mathbf{I}+i\omega\mathbf{U}^\dag\mathbf{H}
\mathbf{U}+\mathbf{D}}$ which couples all modes. Resonant frequencies are ruled by the diagonal matrix $\mathbf{D}$ (the eigenvalues of the system). The size of this set of equations depends on the number of transducers and its modes.

\subsection{The TI Configuration and the Mode Channels}

In order to increase the bandwidth we opted to use six 2-mode mechanical resonators coupled to the antenna surface according to the arrangement suggested by Johnson and Merkowitz: the truncated icosahedron (TI) configuration \cite{johnson1993,merkowitz1997}. This symmetric distribution places transducers in positions which correspond to the centers of pentagonal faces of a TI which minimizes cross-effects. The TI configuration allows us to derive the ``mode channels'' just by the application of the ``pattern matrix'' $\mathbf{B}$.  This matrix contains the numerical value of the five quadrupolar spherical harmonics at each transducer position. So the pattern matrix projects the displacement of each transducer to the normal mode components. In this way the amplitude of the normal quadrupolar modes (the mode channels) can be experimentally monitored by the set of transducers. These mode channel amplitudes are important quantities in data analysis once they are directly related to the tensorial components of the GW as we will show later.

The TI configuration implies Equation \ref{eq:moveq0} to present $17$ second order differential equations (5 from the sphere and 12 from the set of transducers) which can be solved under their symmetry properties. In the TI configuration the last six components of the vector $\mathbf{\tilde{w}}(\omega)$ correspond to the displacement of the last transducer resonator $\mathbf{\tilde{q}}^{R_{2}}(\omega)$ (final mass). We can rewrite these last vector components in the Equation \ref{eq:solmoveq} as
\begin{equation}\label{eq:qR2}
\mathbf{\tilde{q}}^{R_{2}}(\omega)=\mathbf{\tilde{P}}(\omega)\mathbf{\tilde{F}}(\omega),
\end{equation}
where $\mathbf{\tilde{P}}(\omega)$ is the last six lines from the matrix product in Equation \ref{eq:solmoveq}.
When we multiply  $\mathbf{B}$ by Equation \ref{eq:qR2} we obtain an expression for the mode channels which is given by
\begin{equation}\label{eq:gm}
\tilde{\mathbf{g}}(\omega)=\tilde{\mathbf{\xi}}(\omega)\tilde{\mathbf{F}}^S(\omega) +\tilde{\mathbf{\Omega}}^{R_{1}}(\omega)\tilde{\mathbf{F}}^{R_{1}}(\omega)+ \tilde{\mathbf{\Omega}}^{R_{2}}(\omega)\tilde{\mathbf{F}}^{R_{2}}(\omega).
\end{equation}
The quantities denoted by $\tilde{\mathbf{F}}^i(\omega)$ correspond to driven forces which are acting on: the sphere ($S$), the intermediate mass ($R_1$) and the final mass ($R_2$) \cite{costa2004}. The transfer function matrix $\tilde{\mathbf{\xi}}(\omega)$ is the proportionality ratio between the mode channels  and the mode effective forces. We can easily refer the intermediate and final mass noises into the sphere by using the mode response function $\tilde{\mathbf{\Omega}}^{R_{i}}(\omega)$ to the mode $i$ of the transducers. In this way, we projected all noise sources to the antenna modes which is measured as
\begin{equation}\label{eq:modechannels}
\tilde{\mathbf{g}}(\omega)=\tilde{\mathbf{\xi}}(\omega)\tilde{\mathbf{F}}^{\rm{all}}(\omega)
\end{equation}
where $\tilde{\mathbf{F}}^{\rm{all}}(\omega)$ denotes signal plus noise forces which are acting on the modes.

For the moment we can suppose that only a GW signal is exciting the detector and see how this interaction works.

\subsection{The Effective Gravitational Force and the Spherical Amplitudes}

Once a passing gravitational wave generates an acceleration field such effective force (in the frequency domain) is given by \cite{bianchi1996,harry1996,stevenson1997}
\begin{equation}
\tilde{\mathbf{f}}(\omega)=-\frac{1}{2}\omega^2\chi m_SR\tilde{\mathbf{h}}(\omega),
\end{equation}
we can deconvolve the mode channels in the Equation \ref{eq:modechannels} into the spherical amplitudes
\begin{equation}\label{eq:hm}
\tilde{\mathbf{h}}(f)=-\frac{2}{\omega^2\chi m_SR}\tilde{\mathbf{\xi}}^{-1}(\omega)\tilde{\mathbf{g}}(\omega).
\end{equation}
These spherical amplitudes depend on each of the polarization amplitudes and the relative orientation to the lab coordinate system (usually centered in the sphere's center of mass) \cite{merkowitz1998, gasparini2005}. They can be understood as the projected wave polarizations into the sphere's quadrupolar modes under incident angles $\alpha$, $\beta$ e $\gamma$.

\begin{figure}[!t]
\centering
\includegraphics[width=8.4cm]{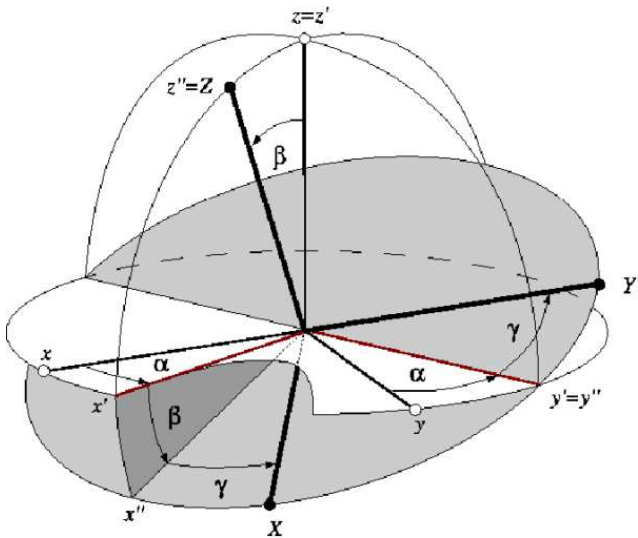}
\caption{The Euler angles which define the relative orientation between $xyz$ and $XYZ$. Rotate $xyz$ counterclockwise around its  axis by $\alpha$ to give $x'y'z'$. Rotate $x'y'z'$ counterclockwise around its $y'$ axis by $\beta$ to give $x''y''z''$. And finally rotate $x''y''z''$ counterclockwise around its $z''$ axis by $\gamma$ to give $XYZ$.\label{fig:yconvention}}
\end{figure}
 The relative orientation between two orthogonal righthanded 3D cartesian coordinate systems $xyz$ (detector frame) and $XYZ$ (wave frame) is described by a $3\times 3$ rotation matrix $\mathbf{R}$, which is commonly parameterized by three so-called Euler angles $\alpha$, $\beta$ e $\gamma$. Figure \ref{fig:yconvention} shows the rotation between these two reference frames. 

We can set the detector frame of such form so that ${y\equiv y'\equiv y''}$ by setting $\alpha=0$, which causes the rotation matrix to be expressed by
\begin{equation}\label{eq:rotmatb}
\mathbf{R}=\left(%
\begin{array}{ccc}
\cos\gamma\cos\beta&\sin\gamma&0\\
-\sin\gamma\cos\beta&\cos\gamma&\sin\gamma\sin\beta\\
\sin\beta&0&\cos\beta
\end{array}%
\right).
\end{equation}
Such an assumption ($\alpha=0$) makes the inverse problem easier to solve considering that only two orientation angles are important: $\beta$ and $\gamma$. These angles easily determine the incoming direction of a candidate signal if it is admitted that $x$-axis matches with local north-south direction and $z$-axis coincides with local zenith. In this case $\beta$ is the complementary angle for the local elevation $\theta$ and $\gamma$ is opposite to the azimuth angle $\phi$. If these angles and the exact arrival time of a signal are known then the sky position of the source can easily be found.

So we can suppose an incident gravitational plane wave (which is carrying the reference frame $XYZ$)  which interacts with the detector (in the reference frame $xyz$) such GW components (in tranverse and traceless form) is represented by
\begin{equation}\label{eq:HTT2b}
\mathbf{H}'(t)\equiv\left(%
\begin{array}{ccc}
  h_{+}(t) & h_{\times}(t) & 0 \\
  h_{\times}(t) & -h_{+}(t) & 0 \\
  0 & 0 & 0 \\
\end{array}%
\right).
\end{equation}
Such components of strain matrix $\mathbf{H}'(t)$ would be projected into a new strain matrix $\mathbf{H}(t)$ by defining $\mathbf{H}(t)=\mathbf{RH}'(t)\mathbf{R}^\dag$.  The projected strain matrix $\mathbf{H}(t)$ presents components as  \cite{merkowitz1998}
\begin{equation}\label{eq:transfmatb}
\mathbf{H}(t)=\left(%
\begin{array}{ccc}
  h_1(t)-\frac{1}{\sqrt{3}}h_5(t) & h_2(t) &  h_4(t)\\
  h_2(t) & -h_1(t)-\frac{1}{\sqrt{3}}h_5(t) &  h_3(t)\\
  h_4(t) & h_3(t) & \frac{2}{\sqrt{3}}h_5(t)\\
\end{array}%
\right).
\end{equation}
These components are linear combinations of the five components of the spherical amplitudes $\mathbf{h}$ presented in Equation \ref{eq:hm} which are denoted by
\begin{subequations}\label{eq:hms}
\begin{equation}
h_1(t)=h_{+}(t)\frac{1}{2}(1+\cos^2\beta)\cos 2\gamma+h_{\times}(t)\cos\beta\sin 2\gamma,\label{eq:h1}
\end{equation}
\begin{equation}
h_2(t)=-h_{+}(t)\frac{1}{2}(1+\cos^2\beta)\sin 2\gamma+h_{\times}(t)\cos\beta\cos 2\gamma,\label{eq:h2}
\end{equation}
\begin{equation}
h_3(t)=-h_{+}(t)\frac{1}{2}\sin 2\beta\sin\gamma+h_{\times}(t)\sin\beta\cos\gamma,\label{eq:h3}
\end{equation}
\begin{equation}
h_4(t)=h_{+}(t)\frac{1}{2}\sin 2\beta\cos\gamma+h_{\times}(t)\sin\beta\sin\gamma,\label{eq:h4}
\end{equation}
\begin{equation}
h_5(t)=h_{+}(t)\frac{1}{2}\sqrt{3}\sin^2\beta.\label{eq:h5}
\end{equation}
\end{subequations}
These equations show us that the spherical amplitudes are extremely dependent on GW polarizations and incident angles. With these assumptions one observes that they also can mimetize both signal and noise depending on what kind of source has been considered.

\subsection{The Noise Sources and the Sensitivity Curve}

Actually as one can see the effective gravitational force in Equations \ref{eq:moveq0} to \ref{eq:modechannels} is not the only force that acts on the sphere. The resultant force $\tilde{\mathbf{F}}^{\rm{all}}(\omega)$ has other components that come from noise sources which are acting on the whole system (e.g. thermal Langevin's forces). Taking into account that we are focused on generating a data time series similar to that which comes from a real detector these sources must be taken into consideration when simulating because they will determine the detector behavior.

Many solutions to minimize the noise sources contributions have been studied and a lot of what has been learned over the last decades with bar instruments has been reused and reorganized to be applied to spherical detectors (e.g. vibration isolation systems, cryogenic devices, electromagnetic isolation, etc). But some noise sources are intrinsic to the instrument and hard to eliminate.

We are interested in the detector sensitivity and its optimum bandwidth when usual noises (thermal, backaction, series and phase noises) \cite{tobar2000b} are acting on it. In order to determine how large the sensitive bandwidth $\Delta f_m$ is  for the mode $m$ we have adopted the relation 
\begin{equation}\label{eq:bandwidth}
\Delta f_m= \frac{1}{2\pi h_{\rm{min}}^2}\int_{-\infty}^{+\infty}{| \tilde h_m \left( \omega  \right)|^2d\omega},
\end{equation} 
where $\tilde h_{\rm{min}}^2$ represents the minimum squared value of $h_m(\omega)$. The knowledge of spectral densities of each of the noise sources makes it possible to determine their contributions to the mode channels and consequently the values of $ \tilde h_m \left( \omega  \right)$.

\begin{figure}[!t]
\centering
\includegraphics[width=8.4cm]{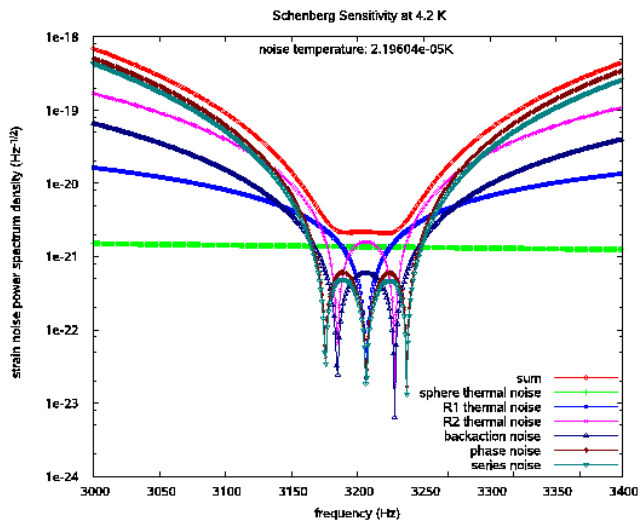}
\caption{\label{fig:sensecurves}The SCHENBERG sensitivity curves at $4.2\,\rm{K}$ (upper  line) for a single quadrupole mode and the individual contributions of the noise sources.}
\end{figure}
Figure \ref{fig:sensecurves} presents the sensitivity curves for a quadrupole mode (all others are considered the same, which they actually are for the degenerated case) and the individual contributions from inputted noise sources. These noise sources are referred into the sphere via transfer function (Equations from \ref{eq:gm} to \ref{eq:hm}).

This fact permits the location of the GW source in the sky, as well as the determination of the polarization amplitudes of the GW.

\subsection{The Signal plus Noise\label{subsec:signalplusnoise}}

To simulate the SCHENBERG behavior when it is excited both by an incident gravitational wave and noise we needed to use some waveform. We adopted the AstroGravS Catalog (\url{http://astrogravs.gsfc.nasa.gov}) where a large number of gravitational waveforms are available. As an example, in this paper we used an adiabatic inspiral of a corotating binary \cite{duez2001}, which was rescaled to double $1.9\,\rm{M}_\odot$ neutron stars (a chirp signal).

 Both signal and noise are expressed in distinct units (i.e. at the frequency domain the signal amplitude is expressed in $\rm{Hz}^{-1}$ and the noise amplitude is expressed in $\rm{Hz}^{-1/2}$), so we turn signal amplitude into another important quantity the characteristic strain which is defined as $h_c(f)=\sqrt{\Delta f |\tilde{h}^{GW}(f)|^2}$, where $|\tilde{h}^{GW}(f)|^2$ is the power spectral density of the signal at frequency $f$ \cite{aufmuth2005}. The characteristic strain is essentially the $rms$ signal in a frequency interval of width $\Delta f$ centered at frequency $f$ and it has same units of noise sources.  If the noise floor is flat then this assumption can be used without restrictions otherwise a correction must be applied. Our correction function has a shape that looks like the ``inverse'' of the sensitivity curve. It assines the right weight for each point of the sensitivity curve. So, the points of maximum sensitivity are multiplied by one and, for example, a point in the curve with sensitivity 100 times worse is multiplied by $1/100$.

After this correction we can introduce signal plus noise to model and estimate the signal-to-noise ratio needed for detection.

\subsection{The Amplitude Signal-to-Noise Ratio ($ASNR$)}

The amplitude signal-to-noise ratio ($ASNR$) was estimated from Thorne (1987) \cite{300years} as
\begin{equation}
ASNR=\sqrt{\int_{-\infty}^{\infty}\tilde{W}(f)\frac{|\tilde{h}^{GW}(f)|^2}{S_N(f)}df}
\end{equation}
where $\tilde{W}(f)$ represents the applied filter function. $S_N(f)$ is the equivalent noise profile. It denotes the sum of the noises which acts on each normal mode. It is given by
\begin{equation}
S_N(f)=\sum_{m=1}^5h_m(f)^2,
\end{equation}
where $h_m(f)$ is the sensitivity of mode $m$ as presented in Figure \ref{fig:sensecurves}. It is easy to show that this sum represents the total noise energy on the sphere. This relation takes in account all sky sensitivity because 
\begin{equation}
\sum_{m=1}^5h_m^2\equiv h^2=h_+^2+h_\times^2.
\end{equation}
In other words it has no incident direction dependence. As we presumed at the beginning the spherical detector really is omnidirectional.

\subsection{The Simulated Data}

In order to generate time series which contain the amplitudes $\mathbf{w}(t)$ and,
consequently, ${\mathbf{q}}_{2}$ (the displacement of the second mechanical mode of the transducers), we introduced first only known noise sources.

We admitted that noise forces behave like the sum of sinusoids with random phases at frequencies $-\infty<\omega<\infty$. However it is impossible to sum an infinite number of sinusoids. The Sampling Theorem guarantees that only frequencies less than the sampling frequency (we use $f_{\mathrm{sample}}=16384\,\mathrm{Hz}$)
could be detected. So we simulated at the frequency domain the sinusoids with amplitudes
$A_i(\omega)=\sqrt{S_i(\omega)}$ that corresponds to spectral densities of noise forces at frequency $\omega$ and we gave them random phases so that
\begin{equation}\label{eq:senoides}
\tilde F(\omega)=A_i(\omega)\left(\cos\varphi+i\sin\varphi\right),
\end{equation}
where $0\leq\varphi<2\pi$ is the chosen phase (with flat distribution) and
$i=\sqrt{-1}$ is the complex unit. The Inverse Fourier
Transform sums the sinusoids to generate the time series. Acting in this way we obtained the noise forces at the time domain. These values are inputted into the model.

\begin{figure}[!ht]
\centering
\includegraphics[height=8.4cm,angle=270]{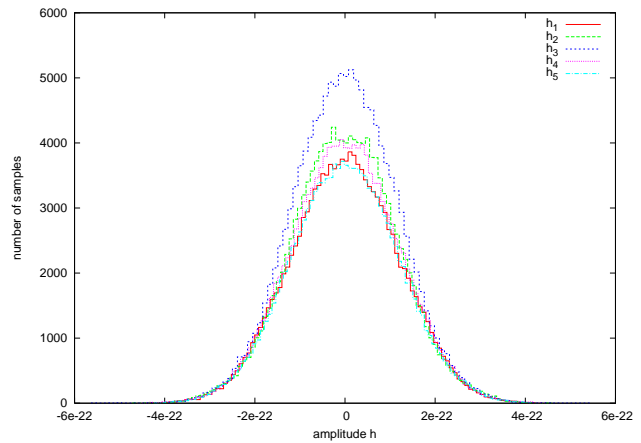}
\caption{Histogram of event values for the five spherical amplitudes $h_m$. Their distribuition is perfectly gaussian for a noise-only case.\label{fig:histhm}}
\end{figure}

The data present a gaussian distribution as expected. With this data to hand we can apply the inverse problem to obtain the spherical amplitudes and find an incident position for each sample. Figure \ref{fig:histhm} presents a histogram of the spherical amplitudes which are obtained from the set of six transducers. The out of band contribution was eliminated by applying a rectangular band-pass filter (the width of such filter is obtained by Equation \ref{eq:bandwidth} and it must be centered in the resonant frequency $f_m$ for the $m$ mode).

From  Figure \ref{fig:histhm} one can also observe that sometimes one mode is more excited ($h_3$ in this collection of data) than another, but the degree of excitation changes from one moment to another which reflects the system dynamics. The energy passes from one mode to another via transducers and also because broken symmetry of the sphere due to the hole machined for the suspension system.

With these quantities we are able to find the ``incident direction'' and the ``GW polarization amplitude'' for each sample by the inverse problem solution.

\section{The inverse problem solution}

In order to obtain ${\mathbf{H}}'(t)$ from ${\mathbf{H}}(t)$ only the angles $\beta$ and $\gamma$ are needed once we adopt the $\alpha=0$ convention. Equation \ref{eq:transfmatb} can be multiplied by ${\mathbf{R}}^\dag$ from left and by ${\mathbf{R}}$ from right and one obtains
\begin{subequations}
\begin{equation}
{\mathbf{R}}^\dag{\mathbf{H}}(t){\mathbf{R}}={\mathbf{R}}^\dag{\mathbf{RH}}'(t){\mathbf{R}}^\dag{\mathbf{R}}.
\end{equation}
Once ${\mathbf{R}}^\dag{\mathbf{R}}={\mathbf{R}}{\mathbf{R}}^\dag={\mathbf{I}}$ we have
\begin{equation}\label{eq:hlabframe}
{\mathbf{H}}'(t)={\mathbf{R}}^\dag{\mathbf{H}}(t){\mathbf{R}}.
\end{equation}
\end{subequations}
The numeric values of $\beta$ and $\gamma$ can be easily obtained by imposing that the last column of ${\mathbf{H}}'(t)$ is null (it really is according to GR, GW has no effect on propagation direction $z'$). Therefore the solutions has the following format
\begin{eqnarray}
\tan\gamma&=&-\frac{y}{x},\label{eq:gamma}\\
\tan\beta&=&\pm\frac{y}{z}\frac{1}{\sin\gamma}\label{eq:beta}.
\end{eqnarray}
The minus sign in Equation \ref{eq:gamma} comes from the Euler y-convention and the double sign in Equation \ref{eq:beta} spots the fact of the sphere is incapable of distinguishing antipodal sources.

The ${\mathbf{H}}'(t)$ determinant is null and the equation system is indeterminate -- one of the equations depends on the others. However only two of them are needed for the univocal determination of $\beta$ and $\gamma$. At this time we impose a contour condition such that when $\tan\gamma=0$ we have $\tan\beta=\infty$, just as it should be \cite{magalhaes1997b}. In this way we have
\begin{subequations}
\begin{equation}
y=h_3(t)h_4(t)-\frac{2\sqrt{3}}{3}h_2(t)h_5(t),
\end{equation}
\begin{equation}
x=-\frac{2\sqrt{3}}{3}\left(h_1(t)h_5(t)+\frac{\sqrt{3}}{3}h_5(t)^2\right)-h_3(t)^2,
\end{equation}
\begin{equation}
z=h_2(t)h_3(t)+h_4(t)\left(h_1(t)+\frac{\sqrt{3}}{3}h_5(t)\right).
\end{equation}
\end{subequations}

In order to obtain an expression for GW polarization we take into account the ${\mathbf{H}}'(t)$ aspect ($h_+=H'_{11}=-H'_{22}$ and $h_\times=H'_{12}=H'_{21}$). One can observe that ${\mathbf{H}}(t)$ and ${\mathbf{H}}'(t)$ are symmetrical so the values of $H'_{12}$ and $H'_{21}$ are always identical and noise independent. However this restriction is not applicable to $H'_{11}$ and $H'_{22}$ so we take $h_+=(H'_{11}+H'_{22})/2$ \cite{merkowitz1998}. So, in this way, we found that
\begin{widetext}
\begin{subequations}
\begin{equation}
h_+(t)=(1+\cos^2\beta)\left(\frac{1}{2}h_1(t)\cos 2\gamma - \frac{1}{2}h_2(t)\sin 2\gamma\right)+
\sin 2\beta\left(-\frac{1}{2}h_3(t)\sin\gamma+\frac{1}{2}h_4(t)\cos\gamma\right)
 +\frac{\sqrt{3}}{2}h_5(t)\sin^2\beta,\label{eq:hmaist}
\end{equation}
\begin{equation}
h_\times(t)= \cos\beta\left(h_1(t)\sin 2\gamma+h_2(t)\cos 2\gamma\right)
+\sin\beta\left(h_3(t)\cos\gamma+h_4(t)\sin\gamma\right)\label{eq:hxt}.
\end{equation}
\end{subequations}
\end{widetext}

These equations can also be derived from Equation \ref{eq:hms} and represent the solution of the inverse problem. They are valid either for noisy and noiseless cases. The noise distribution over the sphere surface can now be estimated for each sample once ``a direction'' is obtained.

\section{The Problem of Incompleteness}

The fact that only the quadrupolar modes have been considered instead of all spherical harmonic functions leads to a problem of incompleteness in the sphere's reconstruction.

\begin{figure}[!ht]
\centering
\includegraphics[width=8.4cm]{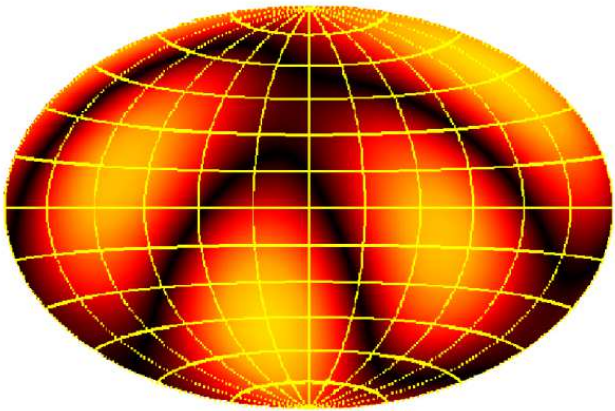}
\caption{The sum of the five quadrupolar modes over the sphere in a Hammer-Aitoff projection (the brightest areas represent the highest absolute amplitudes).\label{fig:summodes}}
\end{figure}
\begin{figure}[!ht]
\centering
\includegraphics[width=8.4cm]{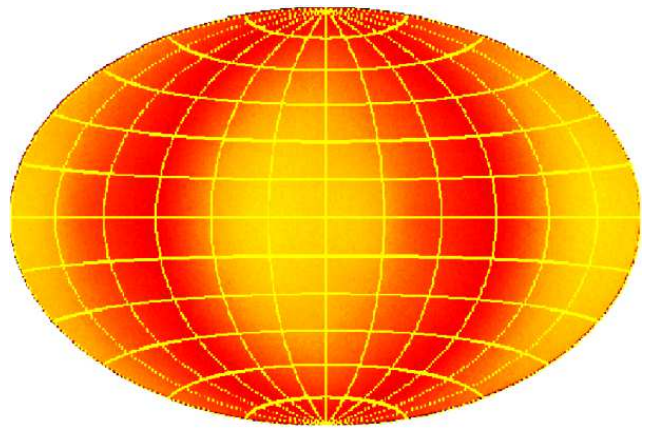}
\caption{When the direction is calculated  the pattern points to some preferential direction. So, the noise isn't uniformly distributed over the sphere.\label{fig:mask}}
\end{figure}

The sum of the five quadrupolar modes (which is shown in Figure \ref{fig:summodes}) with equal degrees of excitation implies in amplified regions which points to a preferential direction (Figure \ref{fig:mask}) when only noise is considered. This direction is actually perpendicular to the position of higher amplitudes of sphere vibrations.

This non-uniform distribution leads us to produce a ``gain mask'' which normalizes the output position map. We call it 
``gain mask'' because either signal or noise that incides in that direction is more easily detected by the inverse solution once it excites all sphere modes equally. An analytic expression for this is not easily obtained so we use a numerical method by using a large number of noise only simulations ($\sim 100$ of $32\,\rm{s}$ at $16384\,\rm{Hz}$) to make it smooth enough to apply it to the problem.

The gain mask enhances areas of low noise levels in the output position map in order to produce an uniform distribution of noise for all points in the map. We introduced a signal such as the one mentioned in Subsection \ref{subsec:signalplusnoise} and located the source at different distances to discover what $ASNR$ is needed to detect it.

\section{The Results}
\begin{figure}[!ht]
\begin{center}
\includegraphics[width=8.0cm]{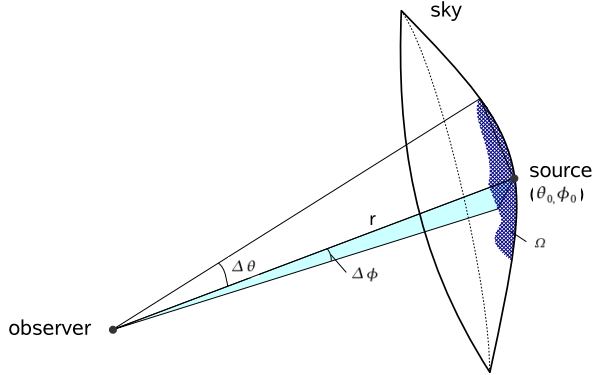}
\end{center}
\caption{The solid angle described by a source if we have a noisy detector. The found position could be inside a solid angle delimited by $\Delta\theta$ and $\Delta\phi$ around its original position ($\theta_0,\phi_0$). Such an area determines the error in position recovering.\label{fig:solidangle}}
\end{figure}

 Once the original source position (in the simulated data) is known we can estimate the precision on its determination. A source must appear at position ($\theta_0,\phi_0$) like a point if no noise is affecting the detector, however in the case of noise it could be found at another place. Low signal-to-noise ratios cause an indetermination in the source location and results in an error. This error puts the source in a place in a region of the sky $\Omega$ (a solid angle) delimited by $\Delta\theta$ and $\Delta\phi$ around its original position. Such an error can be quantified by the solid angle  $\Omega\approx\int_{\theta_0-\Delta\theta}^{\theta_0+\Delta\theta}\int_{\phi_0-\Delta\phi}^{\phi_0+\Delta\phi}\cos\theta d\theta d\phi$. The larger the solid angle filled by the source the larger the error on the position determination is. This effect is presented in Figure \ref{fig:solidangle}.

\begin{figure}[!ht]
\begin{center}
\includegraphics[width=8.4cm]{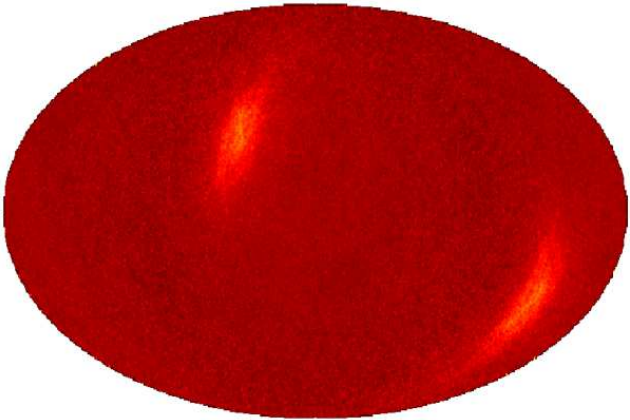}
\end{center}
\caption{\label{fig:sourcelocation} Position of a source with $ASNR\sim 3$. This figure represents the mean of $\sim 150$ distinct simulations.}
\end{figure}
Figure \ref{fig:sourcelocation} shows the incident direction of a source with $ASNR\sim 3$ after we had used a simple band-pass filter on the data. The figure represents the mean amplitude in $\sim 150$ distinct simulations. It is easy to notice that a brighter area appears after these number of simulations. Such an area describes a solid angle that corresponds to the error on the source location. All pointed directions by inverse solution are inside this area. So the area of the source location for $ASNR\sim 3$ case is equivalent $\Omega\approx 0.235\,\rm{sr}$ which approximately corresponds to $2\%$ of the whole sky.

\begin{figure}[!ht]
\begin{center}
\includegraphics[width=8.4cm]{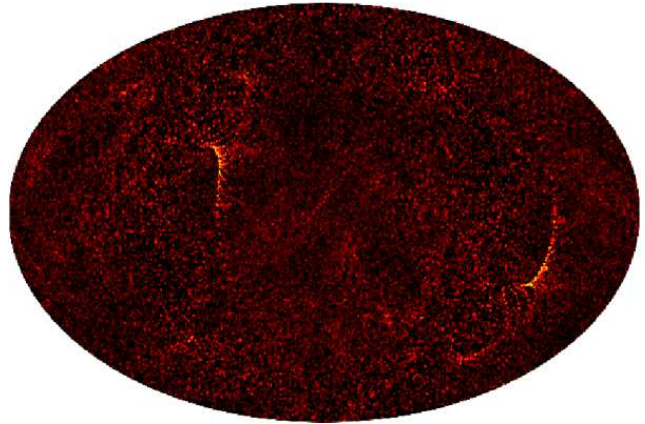}
\end{center}
\caption{\label{fig:sourcelocationb} Position of a source with $ASNR\sim 3$. This figure represents the signal in only one simulation.}
\end{figure}
However, in a single simulation, as shown in Figure \ref{fig:sourcelocationb}, it is already possible to detect the source position. It fills not just a point in the sky but something like a strip during the time that the chirp signal remains in the sensitivity bandwidth. On the other hand, noise samples mark only points randomly distributed over the sphere surface. So behavior like the one showed in Figure \ref{fig:sourcelocationb} is a candidate signature.

\begin{figure}[!ht]
\begin{center}
\includegraphics[height=8.4cm,angle=270]{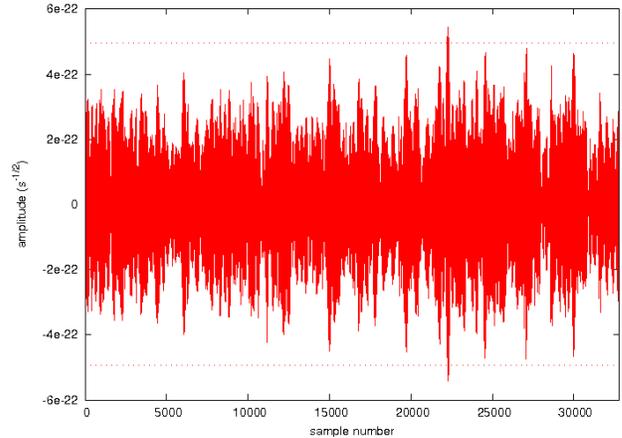}
\end{center}
\caption{\label{fig:hxhplustime} Pseudo-amplitudes from the inverse solution. The dots represents $3\sigma$ of the sample distribution.}
\end{figure}
This $ASNR\sim 3$ represents a trustable limit for a candidate detection. As one can see in Figure \ref{fig:hxhplustime} the level of the signal is almost diving in the noise but its maximum amplitude is $\gtrsim 3\sigma$ which makes the detection possible. The search for signals with lower values for $ASNR$ needs another approach, a statistical one. 

\section{Conclusions}

We numerically solved the inverse problem to recover the information about an input GW source such as its incident direction and polarizations. It is possible to determine event energy and to find a direction for its source even for $ASNR\sim3$. The amazing result emerges from the fact that a spherical detector emulates 5 independent detectors (the mode channels) with an important peculiarity: the simultaneous correlation between them. Neither interpolation nor another artifice is needed.

The mode channels come directly from a product between the pattern matrix and the data and such a computational  operation is easily done in real time. However, the spherical amplitudes (Equation \ref{eq:hm}) depends on the inverse of the transfer function matrix $\tilde{\mathbf{\xi}}(\omega)$ which is difficult to calculate. A solution comes from the fact that such quantity does not depend either on signal or on noise. It only depends on the electro-mechanical features of the detector, so it could be stored on disk or even on the computer memory to be used in a real time analysis. So the presented method can be applied to a real time data analysis software which will be able to emit an on-line warning to be checked.

 This method is very flexible and can be adapted for the real case. Although the model is highly parametrized so the values of the physical parameters must be known. However a set of transfer function with a distinct collection of parameters could be tested offline to validate some candidates.

With this method it becomes evident that a spherical GW detector, like the Mario SCHENBERG, will be able to detect nearby chirp events (up to Magellanic clouds) at its first goal sensitivity ($4.2\,\rm{K}$) just by using a simple band-pass filter.

For $ASNR<3$ (using a simple band-pass filter) another approach must be used. We tested the efficiency of a matched filter and checked that it works well when the waveform and incident direction are known. However it has a poor performance when the input signal doesn't match perfectly with the template. But this will be investigated in detail in a future work.

\begin{acknowledgments}
C.A. Costa would like to thanks to FAPESP for financial support (grant n. 01/14527-3 and 05/00214-4)  and O. D. Aguiar acknowledges FAPESP (grant n. 98/13468-9 and 05/51410-8) and CNPq (grant n. 306467/2003-8) for financial support. The authors also thank Armando Bernui for pointing out the problem of incompleteness.
\end{acknowledgments}

\bibliography{prd06}

\end{document}